\documentclass[preprint]{aastex631}
\usepackage{amsmath}
\usepackage{soul}

\defcitealias{cufari_using_2022}{CCN22}

\received{2024 Octorber 27}
\revised{2024 November 30}
\accepted{2024 December 02}

\shorttitle{Formation rate of quasiperiodic eruptions}
\shortauthors{Cao et al.}

\begin{document}
\title{Formation Rate of Quasiperiodic Eruptions in Galactic Nuclei Containing 
Single and Dual Supermassive Black Holes}

\correspondingauthor{F.K. Liu}
\email{fkliu@pku.edu.cn}

\author[0000-0002-4211-9523]{Chunyang Cao}
\affiliation{Department of Astronomy, School of Physics, Peking University, 
 Beijing 100871, People’s Republic of China}

\author[0000-0002-5310-3084]{F.K. Liu}
\affiliation{Department of Astronomy, School of Physics, Peking University, 
 Beijing 100871, People’s Republic of China}
 \affiliation{Kavli Institute for Astronomy and Astrophysics, Peking University, 
 Beijing 100871, People’s Republic of China} 

\author[0000-0003-3950-9317]{Xian Chen}
\affiliation{Department of Astronomy, School of Physics, Peking University, 
 Beijing 100871, People’s Republic of China}
\affiliation{Kavli Institute for Astronomy and Astrophysics, Peking University, 
 Beijing 100871, People’s Republic of China}

\author[0000-0001-6530-0424]{Shuo Li}
\affiliation{National Astronomical Observatories, Chinese Academy of Sciences, 
Beijing 100012, People's Republic of China}

\begin{abstract}
Quasiperiodic eruptions (QPEs) are a novel class of transients 
recently discovered in a few extragalactic nuclei. 
It has been suggested that a QPE can be produced by 
a main-sequence star undergoing repeated partial disruptions 
by the tidal field of a supermassive black hole (SMBH) 
immediately after getting captured on a tightly bound orbit through the Hills mechanism. 
In this Letter, we investigate the period-dependent formation rate of QPEs for this scenario, 
utilizing scattering experiments and the loss-cone theory. 
We calculate the QPE formation rates in both a single-SMBH and a dual-SMBH system, 
motivated by the overrepresentation of postmerger galaxies as QPE hosts. 
We find that for SMBHs of mass $10^{6}$--$10^{7}M_{\odot}$, 
most QPEs formed in 
this scenario have periods longer than $\simeq 100$ days. 
A single-SMBH system generally 
produces QPEs at a negligible rate of $10^{-10}$--$10^{-8}\ \rm{yr}^{-1}$ 
due to inefficient two-body relaxation. 
Meanwhile, in a dual-SMBH system, 
the QPE rate is enhanced by 3-4 orders of magnitude, 
mainly due to a boosted angular momentum evolution 
under tidal perturbation from the companion SMBH (galaxy). 
The QPE rate in a postmerger galactic nucleus hosting two equal-mass SMBHs 
separated by a few parsecs could reach $10^{-6}$--$10^{-5}\ \rm{yr}^{-1}$. 
Our results suggest that a nonnegligible fraction ($\simeq 10$--$90\%$) of 
long-period QPEs 
should come from postmerger galaxies. 
\end{abstract}


\section{Introduction} \label{sec:introduction}

In the past decade, a growing population of transients has been detected 
in the nuclei of quiescent galaxies by time-domain surveys.  
Many of them show one-off flares lasting a few months to years and 
are considered to be associated with tidal disruption events 
\citep[TDEs;][]{bade_detection_1996,komossa_giant_1999,gezari_ultraviolet_2006,
vanvelzen_opticalultraviolet_2020,gezari_tidal_2021}, 
in which the flare is produced by a supermassive black hole (SMBH) 
tidally disrupting a closely approaching star 
and accreting its debris \citep{hills_possible_1975,rees_tidal_1988}. 
Recently, a distinguishing population of nuclear transients 
has been revealed by observations, 
featuring multiple X-ray and/or optical quasiperiodic eruptions (QPEs) 
with periods ranging from a few or dozens of hours 
\citep{miniutti_ninehour_2019,giustini_xray_2020,
arcodia_xray_2021,arcodia_more_2024,chakraborty_possible_2021,nicholl_quasiperiodic_2024} 
to several weeks \citep{evans_monthly_2023,guolo_xray_2024} 
and up to months 
\citep{payne_asassn14ko_2021,liu_deciphering_2023,liu_rapid_2024}. 
Interestingly, 
the few reported QPEs are preferentially found in postmerger galaxies, 
which are considered to host two SMBHs.  
The QPE host galaxy RXJ1301 \citep{giustini_xray_2020} 
is a postmerger remnant \citep{wevers_xray_2024} 
given its strong poststarburst feature 
and extended reservoirs of gas revealed by emission lines \citep{french_fading_2023}. 
The host galaxy of ASASSN-14ko \citep{payne_asassn14ko_2021} 
shows clear postmerger signatures like 
dual active galactic nucleus (AGN) and a prominent tidal arm \citep{tucker_amusing_2021}. 
The overrepresentation of postmerger galaxies as hosts of QPEs 
possibly implies an enhanced formation rate of QPEs in dual-SMBH systems.

The formation channel of QPEs is still under debate, 
and the viable models could be generally categorized into two groups. 
The first one invokes various accretion-disk instabilities 
\citep{miniutti_ninehour_2019,sniegowska_possible_2020,raj_disk_2021,pan_disk_2022} 
and is mainly applied to GSN$\ 069$, 
the first-ever reported QPE found in an AGN 
(\citealp{miniutti_ninehour_2019}; 
but also see \citealp{pan_application_2023}). 
The second group of models relates the quasiperiodic flares 
to one (or multiple) stellar object orbiting around the central SMBH 
but has different explanations as to how the QPE emissions are powered. 
The proposed scenarios include 
(1) collisions between the stellar object and a preexiting accretion disk 
\citep{sukova_stellar_2021,xian_xray_2021,franchini_quasiperiodic_2023,
linial_emri_2023,tagawa_flares_2023,yao_stardisk_2024}; 
(2) interaction between two stellar extreme mass ratio inspirals 
\citep[EMRIs;][]{metzger_periodic_2017,metzger_interacting_2022}; 
and (3) periodic feeding of the SMBH after the mass 
being stripped from the stellar object over multiple pericenter passages, 
through either Roche Lobe overflows or partial TDEs 
\citep{zalamea_white_2010,campana_multiple_2015,king_gsn_2020,payne_asassn14ko_2021,
krolik_quasiperiodic_2022,wang_model_2022,zhao_quasiperiodic_2022,
lu_quasiperiodic_2023,linial_unstable_2023}.

Based on the partial TDEs scenario, 
\citet*[][hereafter \citetalias{cufari_using_2022}]{cufari_using_2022} 
proposed that the orbiting star should originate from a stellar binary and 
get deposited by Hills capture to a radius vulnerable to partial TDEs, 
namely, the SMBH captures it onto a tightly bound and highly eccentric orbit 
after tidally separating its progenitor stellar binary 
\citep{hills_hypervelocity_1988,ginsburg_fate_2006}. 
They suggest that the \citetalias{cufari_using_2022} scenario can well reproduce 
the observed $\sim 114$ days period of ASASSN-14ko \citep{payne_asassn14ko_2021} 
and provides a general route to produce QPEs of various periods. 
Nonetheless, the formation rate of QPEs and how it varies with the periods are still unclear.  
To answer those questions, 
we provide in this Letter a rate calculation for QPEs of different periods 
in the \citetalias{cufari_using_2022} scenario. 
Motivated by the observational hint of enhanced QPE rate in postmerger galaxies, 
we calculate the rate in both a normal galactic nucleus containing a single SMBH and 
a postmerger nucleus containing dual SMBHs. 

This Letter is organized as follows. 
In Section \ref{sec:methods}, 
we introduce the basic theory and methods to calculate the QPE rate 
in a single-SMBH system (Section \ref{subsec:single-SMBH}) 
and a dual-SMBH system (Section \ref{subsec:double}). 
We then present the results in Section \ref{sec:results} 
and draw our conclusions and make discussions in Section \ref{sec:dis}.

\section{Methods} \label{sec:methods}
In this study, we count the QPEs produced by each Hills-captured star as one event 
regardless of its number of repeating flares.  
To provide meaningful information related to the current and future observation, 
we only consider QPEs of periods shorter than 30 yr ($\approx 10^{4}$ days). 

\subsection{QPE Rate in a Single-SMBH System} \label{subsec:single-SMBH}
We consider a single-SMBH system with a SMBH of mass $M_{\rm{SMBH}}$ 
embedded in the galactic nucleus, 
and the stars around the SMBH are distributed following a spherical double power law: 
\begin{equation} \label{equ:rho_r}
    \rho(r) = 
    \begin{cases}
        \rho_{\mathrm{b}}(\frac{r}{r_{\mathrm{b}}})^{-\gamma}, 
        &r\leq r_{\mathrm{b}} \\
        \rho_{\mathrm{b}}(\frac{r}{r_{\mathrm{b}}})^{-\beta}, 
        &r_{\mathrm{b}}<r<r_{\mathrm{max}} \\
        0, & r\geq r_{\mathrm{max}}. 
    \end{cases}
\end{equation}
Following \citet{liu_enhanced_2013}, 
our fiducial model is set by 
($\gamma=1.75$, $\beta=2$, $r_{\rm{b}}=r_{\rm{i}}$, $r_{\mathrm{max}}=160r_{\rm{i}}$), 
where $r_{\rm{i}}\approx 4.265(M_{\rm{SMBH}}/10^{7}M_{\odot})^{1/2}\ \rm{pc}$ 
is the influence radius of the SMBH 
and $\rho_{\rm{b}}$ is calibrated so that the stellar mass 
enclosed within $r\leq r_{\rm{i}}$ equals to $2M_{\rm{SMBH}}$. 
In the \citetalias{cufari_using_2022} scenario, 
the period-dependent QPE rate is determined by 
the partial TDE rate and the period distribution of the Hills-captured stars. 
The former can be handled analytically with the loss-cone theory 
(Section \ref{subsubsec:rate_bt}), 
and the latter can be studied numerically with scattering experiments 
(Section \ref{subsubsec:period}). 

\subsubsection{Partial TDE Rate of Hills-captured Stars} \label{subsubsec:rate_bt}
The Hills capture rate depends on the flux of stellar binaries 
passing close to the SMBH. 
During a close encounter with the SMBH at a pericenter distance of $r_{\rm{p}}$, 
a circular stellar binary of total mass $m_{\rm{b}}$ and semimajor axis $a_{*}$ 
would be tidally separated with a probability of 
\begin{equation} \label{equ:Pbt}
    P_{\rm{bt}}(r_{\rm{p}})\approx 1-\frac{r_{\rm{p}}}{2.2r_{\rm{bt}}}, 
\end{equation} 
where $r_{\rm{bt}}=(M_{\rm{SMBH}}/m_{\mathrm{b}})^{1/3}a_{*}$ 
is the binary tidal separation radius 
\citep{hills_hypervelocity_1988,bromley_hypervelocity_2006}. 
All the potentially separated binaries with $r_{\rm{p}}\leq 2.2r_{\rm{bt}}$ 
populate a low-angular-momentum region called ``loss cone'' with 
\begin{eqnarray} \label{equ:L_lc}
    L^{2}\leq L_{\rm{lc}}^{2}\approx 2GM_{\rm{SMBH}}(2.2r_{\mathrm{bt}}), 
\end{eqnarray} 
meaning that they would be lost 
from the system due to binary tidal separation, 
binary merger, or full TDE of one or both stellar components 
\citep[e.g.,][]{antonini_tidal_2010,mandel_double_2015}. 
The flux of stellar binaries into the loss cone is mainly driven by two-body scatterings, 
which leads to a diffusion of angular momentum with a speed of 
\begin{eqnarray} \label{equ:L_2}
    \frac{L_{2}^{2}}{T_{\rm{d}}}\simeq \frac{L_{\rm{c}}^{2}}{T_{\rm{r}}} 
\end{eqnarray}
\citep{lightman_distribution_1977}. 
Here $L_{2}$ is the diffusion step length during one 
dynamical timescale: $T_{\rm{d}}\simeq r/\sigma$ 
with $\sigma$ the one-dimensional velocity dispersion calculated from the Jeans equation, 
$L_{\mathrm{c}}\simeq r\sigma$ is the circular angular momentum, and 
\begin{equation} \label{equ:T_r}
    T_{\mathrm{r}}(r)=\frac{\sqrt{2}\sigma^{3}(r)}
    {\pi G^{2}\langle m_{*}\rangle \rho(r)\mathrm{ln}\Lambda}
\end{equation}  
\citep{spitzer_evaporation_1958} 
is the two-body relaxation timescale given the averaged mass of background stars 
$\langle m_{*}\rangle\approx 0.3M_{\odot}$ 
and the Coulomb logarithm 
$\ln{\Lambda}=\ln{(r\sigma^{2}/2G\langle m_{*}\rangle)}\approx 20$ 
\citep{binney_galactic_2011}. 
The differential loss-cone flux at radius $r$ can be assessed with  
\begin{align} \label{equ:R_loss}
    \frac{\mathrm{d}\Gamma_{\rm{lc}}}{\mathrm{d}r} =  
    \frac{\Theta(r)}{T_{\mathrm{d}}(r)}
    \frac{4\pi r^{2}\rho(r)f_{\rm{b}}}{\langle m_{*}\rangle} 
\end{align}
\citep{frank_effects_1976,syer_tidal_1999}, 
where $f_{\rm{b}}$ is the binary frequency and $\Theta$ 
is the fraction of binaries 
\edit1{lost} 
during $T_{\rm{d}}$. 
By analyzing the stellar distribution around the loss cone with the Fokker-Planck equation,  
the fraction is suggested to depend on the strength of two-body relaxation as 
\begin{equation} \label{equ:Theta}
    \Theta = 
    \begin{cases}
        (L_{\rm{2}}^2/L_{\rm{c}}^2) \ln^{-1}{(L_{\rm{c}}^{2}/L_{\rm{lc}}^{2})}, 
        &r\leq r_{\mathrm{cri}} \\
        L_{\rm{lc}}^2/L_{\rm{c}}^2, 
        &r_{\mathrm{cri}}<r<r_{\mathrm{max}}. 
    \end{cases}
\end{equation}
The loss cone is called empty (in ``diffusive'' regime) 
at $r\leq r_{\mathrm{cri}}$ and full (in ``pinhole'' regime) at $r> r_{\mathrm{cri}}$
\citep{lightman_distribution_1977}, 
where $r_{\rm{cri}}$ is the transitional radius between two regimes at which 
$L_{2}^{2}/L_{\rm{lc}}^{2}=\ln{(L_{\rm{c}}^{2}/L_{\rm{lc}}^{2})}$ 
\citep[e.g.,][]{magorrian_rates_1999}.

Among all the loss-cone binaries, 
only a fraction $f_{\rm{Q}}$ of them would end up with QPEs. 
Other binaries are either not tidally separated but consumed by binary mergers or full TDEs 
or are indeed separated but the captured stars are not qualified for partial TDEs. 
To evaluate $f_{\rm{Q}}$, we start by specifying the criterion of partial TDE 
of a star of mass $m_{*}$ and radius $R_{*}$ as 
\begin{equation} 
    \mathcal{R}_{\rm{t}}<r_{\rm{p}}\leq \mathcal{R}_{\rm{pt}}, 
\end{equation}
where $\mathcal{R}_{\rm{t}}$ and $\mathcal{R}_{\rm{pt}}$ are, respectively, 
the physical radius below which the star is fully and partially disrupted. 
The full TDE radius can be expressed as 
$\mathcal{R}_{\rm{t}}=\eta_{\rm{SMBH}}\eta_{\rm{t}}r_{\rm{t}}$ \citep{ryu_tidal_2020} 
based on the typical tidal disruption radius: 
$r_{\rm{t}} = (M_{\rm{SMBH}}/m_{*})^{1/3}R_{*}$.
The first factor 
\begin{equation} \label{equ:dm_SMBH}
    \eta_{\rm{SMBH}}=0.80+0.26\sqrt{\frac{M_{\rm{SMBH}}}{10^{6}M_{\odot}}}
\end{equation} 
accounts for the impacts of relativistic effects \citep{ryu_tidal_2020c}. 
The second factor $\eta_{\rm{t}}$ accounts for various stellar internal structures 
and is suggested to be $0.91$--$1.11$ for low-mass stars ($m_{*}< 0.7M_{\odot}$) and 
$0.37$--$0.67$ for high-mass stars 
\citep[$0.7M_{\odot}\leq m_{*}\leq 1.5M_{\odot}$;][]{law-smith_stellar_2020}. 
The partial TDE radius can be expressed similarly as 
$\mathcal{R}_{\rm{pt}}=\eta_{\rm{SMBH}}\eta_{\rm{pt}}r_{\rm{t}}$. 
According to the hydrodynamical simulations of TDEs, 
the factor $\eta_{\rm{pt}}$ is related to the mass stripped fraction 
during one pericenter passage and can be fit empirically with
\begin{equation} \label{equ:dm_low}
    \frac{\delta{m}}{m_{*}} = 
    \exp{\frac{3.1647-6.3777\eta_{\rm{pt}}^{-1}+3.1797\eta_{\rm{pt}}^{-2}}
    {1-3.4137\eta_{\rm{pt}}^{-1}+2.4616\eta_{\rm{pt}}^{-2}}}, 
    0.5\leq\eta_{\rm{pt}}^{-1} \leq 0.9. 
\end{equation}
\citep{guillochon_hydrodynamical_2013} for low-mass stars and 
\begin{eqnarray} \label{equ:dm_high}
    \frac{\delta{m}}{m_{*}} &=& 
    \left(\frac{\eta_{\rm{t}}}{\eta_{\rm{pt}}}\right)^{\zeta}, \nonumber \\
    \log \zeta &=& 
    0.3+3.15\times 10^{-8}[\log{(M_{\rm{SMBH}}/M_{\odot})}]^{8.42}
\end{eqnarray}
\citep{ryu_tidal_2020,ryu_tidal_2020a,bortolas_partial_2023} 
for high-mass stars. 
Here we consider a partial TDE occurs when $\delta{m}/m_{*}\geq 1\%$ 
\citep[e.g.,][]{broggi_repeating_2024}. 
Solving Equations (\ref{equ:dm_low}) and (\ref{equ:dm_high}) 
combined with Equation (\ref{equ:dm_SMBH}) gives 
$\mathcal{R}_{\rm{pt}}\approx 1.92 (2.94)r_{\rm{t}}$ for low-mass stars 
and $\mathcal{R}_{\rm{pt}}\approx 3.15 (1.99)r_{\rm{t}}$ for high-mass stars 
when $M_{\rm{SMBH}}=10^{6}(10^{7})M_{\odot}$. 
Considering the uncertainties in estimating the full and partial TDE radius 
induced by, e.g., SMBH spins, stellar ages, and stellar models 
\citep[e.g.,][]{kesden_tidaldisruption_2012,law-smith_stellar_2020,sharma_partial_2024}, 
we define the criterion of partial TDEs by a $10^{6}(10^{7})M_{\odot}$ SMBH 
using typical values: 
\begin{eqnarray} \label{equ:pTDE_cri}
    \mathcal{R}_{\rm{t}}=1 (1.6)r_{\rm{t}} < & r_{\rm{p}} &  
    \leq 2 (3)r_{\rm{t}} =\mathcal{R}_{\rm{pt}},\ m_{*}<0.7M_{\odot} \nonumber \\
    \mathcal{R}_{\rm{t}}=0.5 (0.8)r_{\rm{t}} < & r_{\rm{p}} & 
    \leq 3 (2)r_{\rm{t}} =\mathcal{R}_{\rm{pt}},\ m_{*}\geq 0.7M_{\odot}.
\end{eqnarray}
Although the choices of typical values are somewhat arbitrary, 
they would only make a slight difference on the rate calculation 
and are good enough for our pioneer study here.

With the criterion above, $f_{\rm{Q}}$ can be evaluated by counting 
the fraction of Hills-captured stars meeting that requirement. 
Due to insignificant variation of the specific angular momentum during 
a Hills capture, 
$r_{\rm{p}}$ is approximately the same as that of 
the injection orbit of the stellar binary \citep[e.g.][]{generozov_hills_2020} 
and thus depends on the type of the loss-cone regime. 
In the pinhole regime, a stellar binary has a one-off close interaction with the SMBH 
at uniformly distributed $r_{\rm{p}}$.
Combined with the binary separation probability (see Equation (\ref{equ:Pbt})) 
and a binary merger fraction of $\simeq 6\%$ 
\citep[e.g.,][]{bromley_hypervelocity_2006,mandel_double_2015}, we have 
\begin{equation} \label{equ:f_full}
    f_{\rm{Q}}\approx 
    (1-6\%)\times 
    \frac{\int_{\mathcal{R}_{\rm{t}}}^{\min({\mathcal{R}_{\rm{pt}},2.2r_{\rm{bt}}})}
    P_{\rm{bt}}(r_{\rm{p}})\rm{d}r_{\rm{p}}}
    {\int_{0}^{2.2r_{\rm{bt}}}\rm{d}r_{\rm{p}}}.
\end{equation}
In the diffusive regime, 
a stellar binary could have multiple close interactions with the SMBH. 
The typical number of close interactions is approximately $L_{\rm{lc}}^{2}/L_{2}^{2}$. 
Therefore, most (say $99\%$) binaries would have been tidally separated 
before reaching $r_{\rm{e}}$, 
where $1-[1-P_{\rm{bt}}(r_{\rm{e}})]^{L_{\rm{lc}}^{2}/L_{2}^{2}}\simeq 99\%$. 
We assume that $r_{\rm{p}}$ is uniformly distributed between 
$r_{\rm{e}}$ and $2.2r_{\rm{bt}}$. 
Moreover, considering that about $75\%$ of binaries would merge 
in the empty loss cone \citep{bradnick_stellar_2017}, we have 
\begin{eqnarray} \label{equ:f_empty}
    f_{\rm{Q}}
    \approx (1-75\%)\times 
    \begin{cases}
        0, \ &r_{\rm{e}} \geq  \mathcal{R}_{\rm{pt}} \\
        \frac{\min{(\mathcal{R}_{\rm{pt}},2.2r_{\rm{bt}})} - 
        \max{(r_{\rm{e}}, \mathcal{R}_{\rm{t}})}}
        {2.2r_{\rm{bt}}-r_{\rm{e}}}, \   &r_{\rm{e}} < \mathcal{R}_{\rm{pt}}.
    \end{cases}
\end{eqnarray}

\subsubsection{Period Distribution of Hills-captured Stars} \label{subsubsec:period}
The calculations above do not involve the periods of QPEs. 
Theoretically, the QPE period equals to the orbital period of the Hills-captured star: 
\begin{eqnarray} \label{equ:Pcap}
    P_{\rm{c}}&=&
    2\pi\left(\frac{a_{\rm{c}}^{3}}{GM_{\rm{SMBH}}}\right)^{1/2} \nonumber \\ 
    &\approx & 15.8 \ \rm{yr} \left(\frac{M_{\rm{SMBH}}}{10^{6}\ M_{\odot}}\right)^{1/2}
    \left(\frac{m_{\rm{e}}}{M_{\odot}}\right)^{-3/2}
    \left(\frac{m_{\rm{e}}+m_{\rm{c}}}{2\ M_{\odot}}\right)^{1/2}
    \left(\frac{a_{*}}{0.1\ \rm{AU}}\right)^{3/2},
\end{eqnarray}
where $m_{\rm{c}}$ and $m_{\rm{e}}$ are the mass of the captured and the ejected stars, 
respectively, and $a_{\rm{c}}=Gm_{\rm{c}}M_{\rm{SMBH}}/(2\Delta{E})$ 
is the semimajor axis of the capture orbit provided 
an energy exchange of 
\begin{equation} \label{equ:DeltaE}
    \Delta{E}\simeq  \frac{Gm_{\rm{c}}m_{\rm{e}}}{a_{*}}
    \left(\frac{M_{\rm{SMBH}}}{m_{\rm{c}}+m_{\rm{e}}}\right)^{1/3}
\end{equation}
\citep{hills_hypervelocity_1988} during the binary tidal separation. 
However, the real energy exchange and the related orbital period 
could vary from those typical values 
by up to 1 order of magnitude depending on the ratio $r_{\rm{p}}/r_{\rm{bt}}$, 
the orbital direction, and the phase of the stellar binary   
\citep[e.g.,][]{sari_hypervelocity_2010,rossi_velocity_2014}. 
Therefore, to obtain a physical distribution of $P_{\rm{c}}$, 
we perform scattering experiments of stellar binaries tidally separated 
in a single-SMBH system.

The properties of the stellar binaries in the scattering experiments 
are summarized in Table \ref{tab:para}, and we cover them briefly here. 
The mass of the stellar binary is fixed at $m_{1}=m_{2}=1M_{\odot}$, 
and the scattering experiences' results could be scaled to binaries of 
various masses based on Equation (\ref{equ:Pcap}). 
The semimajor axis range of $0.005\leq a_{*}\leq 0.2\ \rm{au}$ 
covers the potential progenitor binaries of QPEs, 
as binaries of larger $a_{*}$ hardly produce QPEs of periods $\leq 10^{4}\ \rm{days}$ 
(see Equation (\ref{equ:Pcap})), 
and those of smaller $a_{*}$ suffer severely from binary mergers. 
The rotation of two stars in the binary is of isotropic direction and random phase 
as realized by the three angles ($\theta_{*},\omega_{*},\phi_{*}$). 

We perform $10^{6}$ scattering experiments for each $a_{*}$ and $M_{\rm{SMBH}}$. 
In each experiment, 
the stellar binary is launched from the influence radius 
on a parabolic orbit around the SMBH with an orbital pericenter $r_{\rm{p}}$ 
uniformly sampled between $[0,\ 5r_{\rm{bt}}]$ (mimicking a full loss cone). 
Its orbit is numerically integrated with the program 
Fewbody \citep{fregeau_stellar_2004}, 
where we set both the absolute and relative accuracy to $10^{-11}$. 
The integration is terminated when: 
\begin{itemize}
    \item The binary escapes from the SMBH to 
    $r>50r_{\mathrm{i}}$ with nonnegative energy while keeping self-bounded.
    \item The binary is tidally separated and the ejected star reaches $r>r_{\mathrm{i}}$. 
    \item The CPU integration time exceeds $1200\ \rm{s}$. 
\end{itemize}
The last condition is adopted for computation efficiency, 
and we have checked that only a small fraction ($\lesssim 0.6\%$)
of simulations is subjected to it.
We record the minimum separation between the two stars: $R_{\rm{min*}}$ 
and those between either star and the SMBH: $R_{\rm{min1}}$, $R_{\rm{min2}}$. 
For every Hills-captured star, 
we calculate its period ($P_{\rm{c}}$) and pericenter distance ($r_{\rm{c}}$). 

\begin{deluxetable*}{cccc}
    \tablenum{1}
    \tablecaption{Summary of parameters in the scattering experiments \label{tab:para}}
    \tablewidth{0pt}
    \tablehead{
    \colhead{System} & \colhead{Description} & \colhead{Parameter} & \colhead{Value or Distribution} 
    }
    \startdata
    & mass of the primary star & $m_{1} (M_{\rm{\odot}})$  & $1$ \\ 
    & mass ratio & $q_{*} $  & $1$ \\ 
    Stellar & semimajor axis & $a_{*}$ (au) &  $0.005,0.0075,0.01,0.015,0.025,0.0035,0.05,0.1,0.2$ \\
    Binaries & eccentricity & $e_{*}$ &   thermally distributed between $[0,\ 1]$  \\ 
    & inclination & $\theta_{*}$ & $\cos{\theta_{*}}$ uniformly distributed between $[-1,\ 1]$ \\
    & longitude of ascending node & $\omega_{*}$ & uniformly distributed between $[0,\ 2\pi]$ \\
    & initial phase & $\phi_{*}$ & uniformly distributed between $[0,\ 2\pi]$ \\
    \hline
    Single- & mass of the SMBH & $M_{\rm{SMBH}} (M_{\rm{\odot}})$  & $10^{6},\ 10^{7}$ \\ 
    SMBH & pericenter distance of the injecting binaries & $r_{\rm{p}}$ & uniformly distributed between $[0,\ 5r_{\rm{bt}}]$ \\ 
    \hline 
    & mass of the primary SMBH & $M_{1} (M_{\rm{\odot}})$  &  $10^{6},\ 10^{7}$ \\  
    & mass ratio & $q_{\rm{SMBH}}$ & 1 \\
    Dual- & eccentricity & $e_{\rm{SMBH}}$ & 0 \\
    SMBH & separation between two SMBHs & $d_{\rm{SMBH}}$ & $2r_{\rm{i}}$ \\
    & initial radius of binaries to the primary SMBH & $r$ & $0.1r_{\rm{i}}$--$2d_{\rm{SMBH}}$ 
    \enddata
    \tablecomments{The first section shows the properties of the stellar binaries. 
    The second (third) section is the setup of the single-SMBH (dual-SMBH) system. 
    }
\end{deluxetable*}

From the scattering experiments, 
we can obtain the empirical cumulative distribution function of $P_{\rm{c}}$: 
$F_{\rm{P}}$ for a given progenitor stellar binary with the following steps.  
First, we select successfully Hills-captured stars, 
requiring $R_{\rm{min*}}>R_{\rm{1}}+R_{\rm{2}}$ to exclude merged binaries 
and $R_{\rm{min1(2)}}>\mathcal{R}_{\rm{t1}(2)}$ to exclude full TDEs, 
where $R_{\rm{1(2)}}$ and $\mathcal{R}_{\mathrm{t1(2)}}$ are, respectively, 
the stellar radius and the full TDE radius of the two stars. 
Next, we evaluate $F_{\rm{P}}$ in the full (empty) loss cone by counting 
all the successfully Hills-captured stars (only those of $r_{\rm{c}}\geq r_{\rm{e}}$). 
Finally, we interpolate $F_{\rm{P}}$ as a linear function of $r_{\rm{bt}}$ 
and scale it according to Equation (\ref{equ:Pcap}) to 
account for varying semimajor axis and stellar mass.

\subsubsection{Mock Binary Population} \label{subsec:rate_mock}
We calculate the QPE rate for a mock population of stellar binaries, 
generated by Monte Carlo sampling 
of their primary stellar mass $m_{1}$, mass ratio $q_{*}$ and semimajor axis $a_{*}$. 
We sample $m_{1}$ following the Kroupa initial mass function 
\citep[IMF;][]{kroupa_variation_2001} 
in the range of $0.08M_{\odot}$--$120M_{\odot}$. 
We assume the binary population is $3\times 10^{9}\ \rm{yr}$ old. 
Moreover, according to the main sequence (MS) lifetime, 
\footnote{All the stellar properties in this study are calculated 
with analytic formulas in \citet{hurley_comprehensive_2000} given metallicity $Z=0.02$.} 
about $94.8\%$ of them with $m_{1}\leq 1.45 M_{\odot}$ are still in MS, 
while more massive ones would have evolved into compact objects (COs), 
including $\sim 4.56\%$ white dwarfs (WDs) with $1.45M_{\odot}<m_{1}\leq 8 M_{\odot}$, 
$\sim 0.39\%$ neutron stars (NSs) with $8M_{\odot}<m_{1}\leq 20 M_{\odot}$, 
and $\sim 0.15\%$ stellar-mass black holes (sBHs) with $m_{1}>20 M_{\odot}$
\citep{belczynski_compact_2008}. 
For MS-MS binaries, we sample their properties according to \citet{offner_origin_2022}, 
including an increasing binary frequency $f_{\rm{b}}$ and log-normal 
distributions of $a_{*}$ as functions of $m_{1}$, 
and a flat distribution of $q_{*}$. 
For CO-MS binaries, we set $f_{\rm{b}}=$ 0.3 (WD-MS), 0.07 (NS-MS), and 0.1 (sBH-MS) 
following \citet{antonini_secular_2012}. 
The COs' mass and the binary periods are sampled according to 
either the binary evolution synthesis results 
\citep[for WD-MS and sBH-MS;][]
{belczynski_comprehensive_2004,willems_detached_2004,davis_comprehensive_2010} 
or the pulsar binary observation 
\citep[for NS-MS;][]{manchester_atnf_2005,lattimer_nuclear_2012}, 
and the masses of the MS companions are sampled from 
$0.08M_{\odot}$ to $1.45M_{\odot}$ following the Kroupa IMF. 

For each mock stellar binary, 
we calculate its specific QPE rate ($\Gamma_{\rm{j}}$). 
The captured star is randomly chosen from two stellar components,  
as it is concluded that they have equal probabilities of capture 
regardless of $q_{*}$ \citep{sari_hypervelocity_2010,kobayashi_ejection_2012}. 
We have performed additional sets of scattering experiments with $q_{*}=(0.1, 0.25, 0.5)$ 
and arrived at the same conclusion. 
Note that the conclusion above is only valid when the binaries are compact 
(like those of interest here) or approach the SMBH 
on parabolic orbits with initial energy $E_{0}=0$, 
so that the final energy of two stars is dominated by the energy exchange 
during the tidal separation (see Equation (\ref{equ:DeltaE})) rather than $E_{0}$. 
Otherwise, the heavier (lighter) star is more likely to be captured 
after the tidal separation of a loose binary on a elliptic (hyperbolic) orbit 
\citep[see][]{miller_binary_2005,bromley_hypervelocity_2006,kobayashi_ejection_2012}. 
If a CO is captured, we directly assign $\Gamma_{\rm{j}}=0$. 
Otherwise, $\Gamma_{\rm{j}}$ is calculated with (see Equation (\ref{equ:R_loss})) 
\begin{align} \label{equ:Gamma_single}
    \Gamma_{\rm{j}}(P) = 
    \int_{r_{\rm{min}}}^{r_{\rm{max}}} f_{\rm{Q}}F_{\rm{P}}(P)
    \frac{\Theta}{T_{\mathrm{d}}}
    f_{\rm{b}}\frac{4\pi r^{2}\rho}{\langle m_{*}\rangle}\mathrm{d}r, 
\end{align}
where $F_{\rm{P}}(P)$ is evaluated from our scattering experiments, 
as introduced in Section \ref{subsubsec:period}, 
and $r_{\rm{min}}$ is the minimum radius where the stellar binary keeps hard 
against the ionization effect of local field stars, 
i.e., its self-binding energy is larger than 
$\langle m_{*}\rangle \sigma^{2}(r)$ \citep{binney_galactic_2011}. 
The QPE rate in a single-SMBH system is obtained by averaging of $\Gamma_{\rm{j}}$ 
calculated for $10^{7}$ mock binaries of each type.

\subsection{QPE rate in a Dual-SMBH system} \label{subsec:double}
In a dual-SMBH system, stellar binaries initially orbiting around one SMBH would evolve 
differently under the perturbation by the other SMBH (galaxy).
We investigate the impact of that perturbation on the QPE rate 
by performing scattering experiments of a stellar binary tidally separated 
in a dual-SMBH system.

In the scattering experiments, 
two equal-mass SMBHs are separated by $d_{\rm{SMBH}}=2r_{\rm{i}}$ 
and rotate around the coordinate origin on a circular orbit 
in the combined gravitational field of the two SMBHs and their host galaxies 
based on the density profile of Equation (\ref{equ:rho_r}). 
The stellar binaries share the same properties introduced above (see Table \ref{tab:para}). 
Moreover, the scattering experiments are performed in mostly the same way 
as those for a single-SMBH system, except that: 
\begin{enumerate}
    \item Initially, the stellar binary is placed isotropically around the primary SMBH 
    at radius between $0.1r_{\rm{i}}$ and $2d_{\rm{SMBH}}$. 
    For an unbound binary at $r\geq r_{\rm{i}}$, 
    its initial velocities relative to the primary SMBH 
    (in each of three directions) are sampled from 
    a gaussian distribution with dispersion $\sigma$. 
    For a bound binary at $r< r_{\rm{i}}$, 
    we launch it from the apocenter of its Kepler-like orbit around the SMBH 
    with semimajor axis $a=r$ and thermally distributed eccentricity. 
    \item The orbit is integrated for $5P_{\rm{SMBH}}$ with $P_{\rm{SMBH}}$
    the orbital period of the dual-SMBH system, 
    beyond which the two SMBHs would spiral in significantly due to dynamical friction 
    (see Equation (\ref{equ:Tdf})). 
    The Hills capture events in the first $P_{\rm{SMBH}}$ are discarded 
    to eliminate the impact of initial conditions. 
    \item The number of scattering experiments for each $a_{*}$ and $M_{\rm{SMBH}}$ 
    is increased to $10^{8}$.
\end{enumerate}

Then for a mock stellar binary, its specific QPE rate is calculated with 
\begin{equation} \label{equ:Gamma_dual}
    \Gamma_{\rm{d,j}}(P) = 2\int_{\max{(r_{\rm{min}},0.1r_{\rm{i}})}}^{2d_{\rm{SMBH}}} 
    \left[\rm{F}_{\rm{r}}(\mathcal{R}_{\rm{pt}})-\rm{F}_{\rm{r}}(\mathcal{R}_{\rm{t}})\right]
    \rm{F}_{\rm{P}}(P) 
    f_{\rm{b}}
    \frac{\mathrm{d}\Gamma_{\rm{bt}}}{\mathrm{d}r}
    \mathrm{d}r,
\end{equation} 
where the factor $2$ accounts for QPEs by two SMBHs, 
$\Gamma_{\rm{bt}}$ is the Hills capture rate,  
and $\rm{F}_{\rm{r}}$ and $\rm{F}_{\rm{P}}$ are the empirical cumulative distribution 
function of $r_{\rm{c}}$ and $P_{\rm{c}}$ respectively. 
The parameters $\Gamma_{\rm{bt}}$ and $F_{\rm{r}}$ are both 
evaluated from successfully Hills-captured stars 
(see requirements in Section \ref{subsubsec:period}) 
and linearly interpolating as a function of $r_{\rm{bt}}$. 
Moreover, $F_{\rm{P}}$ is evaluated with a further step of 
scaling $P_{\rm{c}}$ as a function of $m_{\rm{c}}$ and $m_{\rm{e}}$ 
according to Equation (\ref{equ:Pcap}).

\section{Results} \label{sec:results}

\begin{figure*}
    \centering 
    \includegraphics[width=0.94\textwidth]{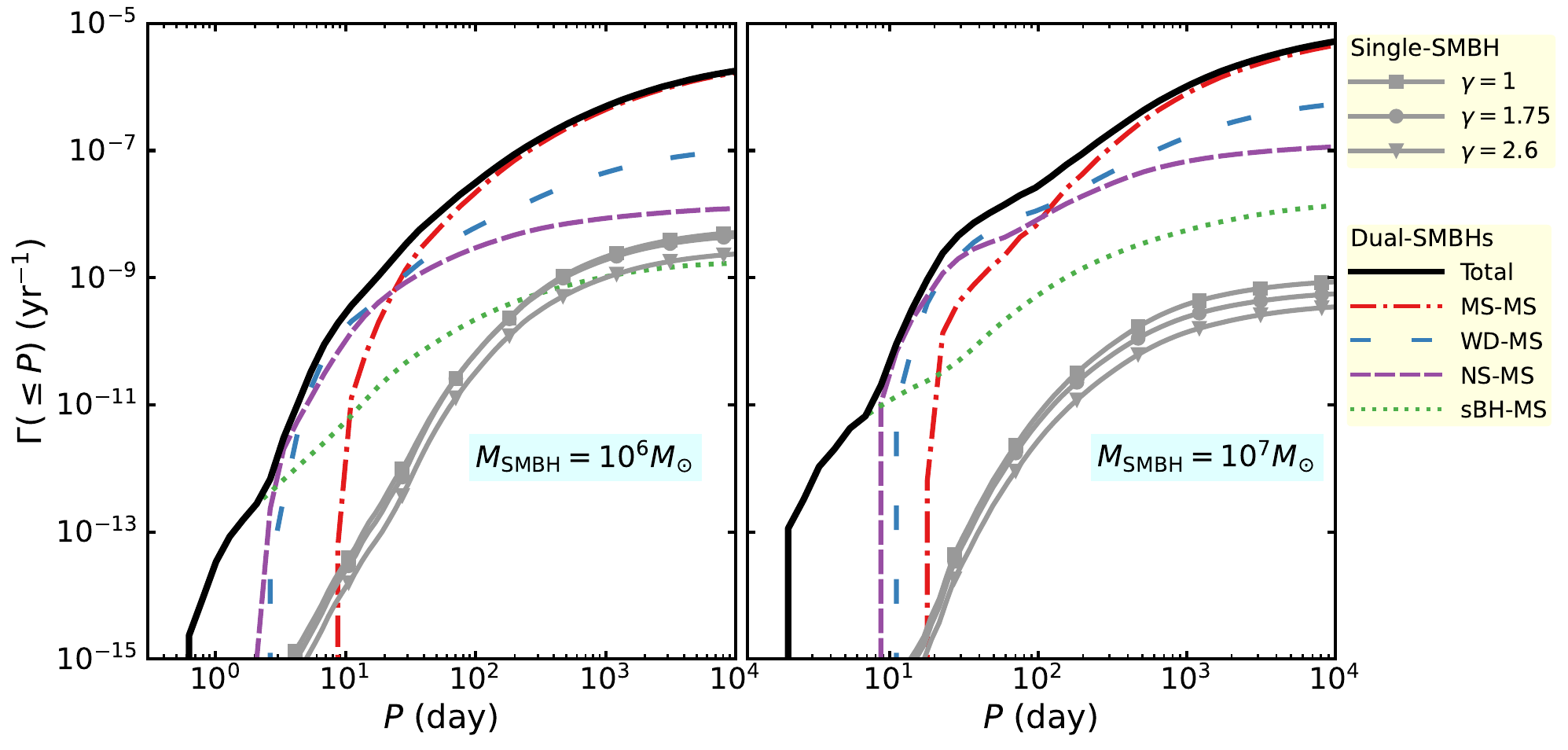}
    \caption{Rate of QPEs ($\Gamma$) produced by Hills-captured stars 
    with periods shorter than $P$ when the SMBH is of $10^{6}M_{\odot}$ (left panel) 
    and $10^{7}M_{\odot}$ (right panel). 
    The gray lines show the rates in a single-SMBH system where stellar binaries are 
    scattered to the SMBH by two-body relaxation in the galactic nucleus of 
    cored ($\gamma=1$, squared), cuspy ($\gamma=1.75$, circle), 
    and extremely cuspy ($\gamma=2.6$, triangle) density profiles. 
    The black lines show the rates in a dual-SMBH system consisting of 
    two equal-mass SMBHs separated by $2r_{\rm{i}}$, 
    and the contributions from four subgroups of progenitor stellar binaries 
    are separately shown: 
    MS-MS (red dashdot), WD-MS (blue loosely dashed), NS-MS (purple densely dashed), 
    and sBH-MS (green dotted).}
    \label{fig:period_dis}
\end{figure*}

We calculate the formation rate of QPEs in a single-SMBH and a dual-SMBH system, 
provided that they are produced by the \citetalias{cufari_using_2022} scenario, 
i.e., repeated partial TDEs of a Hills-captured star immediately following 
the tidal separation of its progenitor stellar binary. 
Our results are presented in Figure \ref{fig:period_dis}, which shows 
the cumulative rate of QPEs with periods shorter than a specific $P$, 
and the QPE rates in different period ranges are explicitly listed in Table \ref{tab:rate}. 

\begin{figure}
    \centering
    \includegraphics[width=0.94\textwidth]{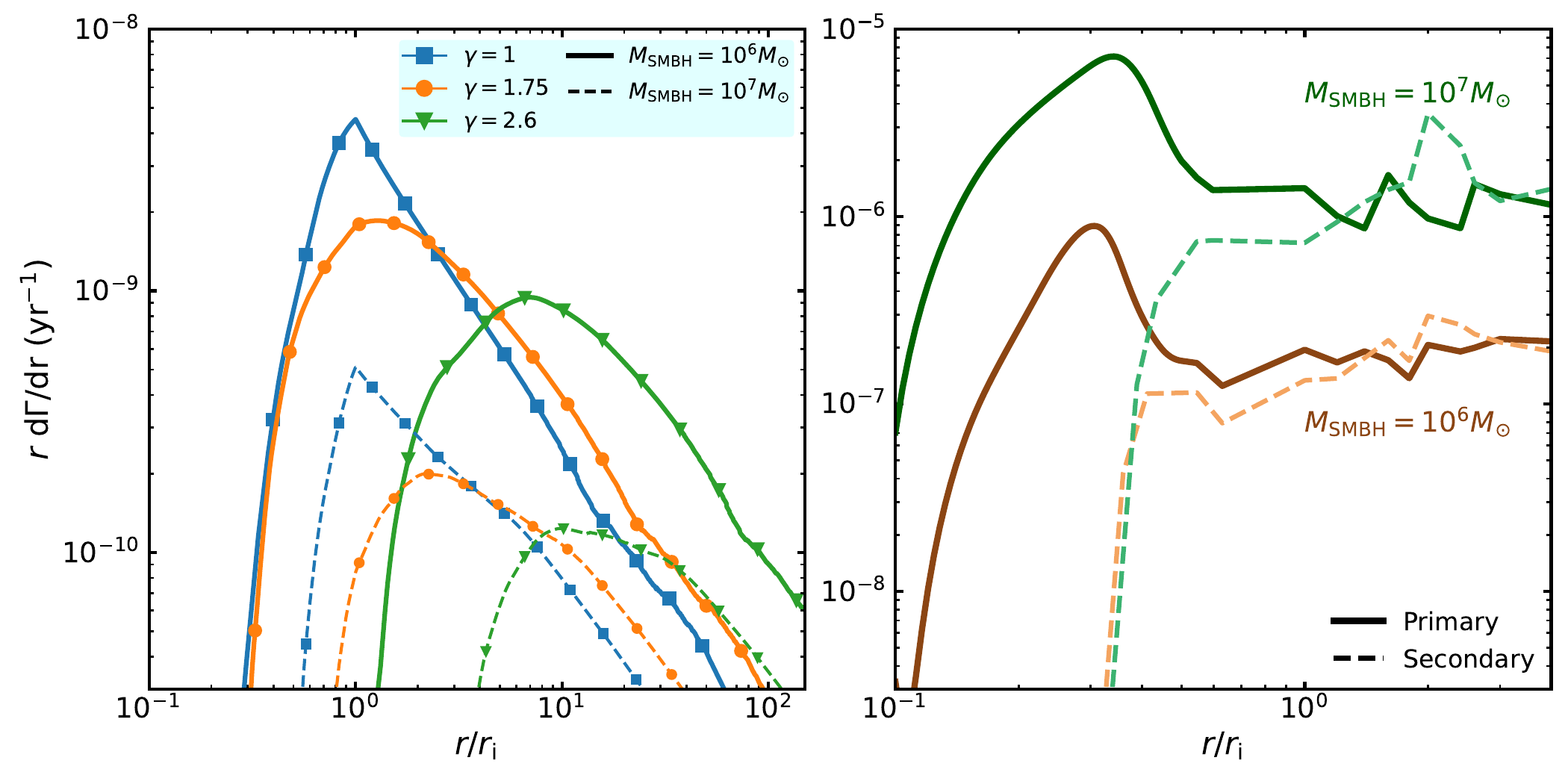}
    \caption{
    Formation rate of QPEs of periods $\leq 10^{4}\ \rm{days}$ 
    contributed by binaries at different distances 
    to the SMBH in unit of the influence radius of the SMBH $r_{\rm{i}}$. 
    The left panel is for a single-SMBH system of 
    different inner stellar density profiles--- 
    cored ($\gamma=1$, blue squared), cuspy ($\gamma=1.75$, orange circle), 
    and extremely cuspy ($\gamma=2.6$, green triangle) 
    and of two SMBH masses---$10^{6}M_{\odot}$ (thick solid) and $10^{7}M_{\odot}$ (thin dashed). 
    The right panel is for an equal-mass dual-SMBH system of SMBH mass 
    $10^{6}M_{\odot}$ (brown) and $10^{7}M_{\odot}$ (green), 
    where the distance is relative to the primary SMBH and 
    the secondary SMBH is located at $r=2r_{\rm{i}}$. 
    The solid (dashed) lines show the contributions of QPEs from the primary (secondary) SMBH.} 
    \label{fig:rate_dr}
\end{figure}

In a single-SMBH system, the QPE rate is of order $10^{-10}$--$10^{-8}\ \rm{yr}^{-1}$. 
It is about 4--5 orders of magnitude lower than the (repeated) partial TDE rate 
from single field stars \citep[e.g.,][]{bortolas_partial_2023}, 
primarily owing to the scarcity of binary stars compact enough to produce QPEs of periods 
$\leq 10^{4}\ \rm{days}$. 
The QPE rate is higher in a less massive SMBH system, 
which is expected as two-body relaxation is more efficient in scattering binaries 
to the SMBH with decreasing SMBH mass \citep[e.g.,][]{wang_revised_2004}. 
Apart from the SMBH mass, 
the two-body relaxation efficiency (and hence the loss-cone flux) 
is also suggested to vary with the stellar distribution 
\citep[e.g.,][]{magorrian_rates_1999,stone_enhanced_2016}. 
To explore the impact of the stellar distribution on the QPE rate, 
we adopt different values of $\gamma$ in Equation (\ref{equ:rho_r}), 
representing cored ($\gamma=1$), cuspy ($\gamma=1.75$), 
and extremely cuspy ($\gamma=2.6$, in reference to the centrally overdense E+A galaxy 
studied in \citealp{stone_enhanced_2016}) inner density profiles. 
Our results suggest that the QPE rate is insensitive to $\gamma$. 
To better understand this feature, we show in Figure \ref{fig:rate_dr}
the QPE rate contributed by progenitor binaries from different initial radius to the SMBH. 
Compared to the empty loss-cone flux \citep[e.g.,][]{wang_revised_2004},
the QPE rate is clearly suppressed at $r\leq r_{\rm{i}}$ 
because the binary fraction drops quickly therein due to ionization effect 
\citep[e.g.,][]{hopman_binary_2009}, 
so it is not surprising that 
the impact of inner density profiles is effectively erased.

In a dual-SMBH system, the total QPE rate is $10^{-6}$--$10^{-5} \rm{yr}^{-1}$, 
about 3--4 orders of magnitude higher than that in a single-SMBH system. 
The rate increases with the SMBH mass, just contrary to the single-SMBH case. 
Most QPEs are contributed by stellar binaries 
initially bound to the primary SMBH at about $0.3r_{\rm{i}}$, 
as shown in Figure \ref{fig:rate_dr} 
(depending on the separation of two SMBHs). 
The rate enhancement is attributed to the tidal perturbation from the secondary SMBH (galaxy), 
which evolves the angular momentum of stellar binaries more efficiently 
compared to the two-body scatterings in a single-SMBH system. 
The efficient angular momentum evolution favors the formation of QPEs in three aspects.  
First, it directly leads to a larger flux of stellar binaries to the vicinity of the SMBH 
\citep{roos_galaxy_1981} and eventually a higher Hills capture rate.  
Second, the stellar binaries can be brought to and get tidally separated at pericenters 
closer to the SMBH, 
so that the Hills-captured stars are more vulnerable to partial TDEs.  
Lastly, the stellar binaries are preferentially tidally separated rather than merged 
during their one-time strong tidal interaction with the SMBH \citep{mandel_double_2015}. 
By contrast, binaries in a single-SMBH system are more likely to merge 
because their eccentricities could be excited by tidal interactions 
during multiple close pericenter passages \citep{bradnick_stellar_2017}.

\begin{deluxetable*}{ccDDDDDD}
    \tablenum{2}
    \tablecaption{Formation Rate of QPEs from Hills-captured Stars \label{tab:rate}}
    \tablewidth{0pt}
    \tablehead{
    \colhead{System} & \colhead{$M_{\rm{SMBH}}$} & \twocolhead{$\gamma$} & \twocolhead{$\Gamma(\leq 10\ \rm{day})$} & 
    \twocolhead{$\Gamma(10-10^{2}\ \rm{day})$} & \twocolhead{$\Gamma(10^{2}-10^{3}\ \rm{day})$} & 
    \twocolhead{$\Gamma(10^{3}-10^{4}\ \rm{day})$} & \twocolhead{$\Gamma(\leq 10^{4}\ \rm{day})$} \\ 
    \colhead{...} & \colhead{$(M_{\odot})$} & \twocolhead{...} & \twocolhead{$(\rm{yr}^{-1})$} & \twocolhead{$(\rm{yr}^{-1})$} & 
    \twocolhead{$(\rm{yr}^{-1})$} & \twocolhead{$(\rm{yr}^{-1})$} & \twocolhead{$(\rm{yr}^{-1})$}
    }
    \decimalcolnumbers
    \startdata
    {       } & $10^{6}$ & $1.0$  & $3.3\times 10^{-14}$ & $6.4\times 10^{-11}$ & $2.1\times 10^{-9}$  & $3.0\times 10^{-9}$  & $5.1\times 10^{-9}$  \\
    {       } & $10^{6}$ & $1.75$ & $2.5\times 10^{-14}$ & $6.4\times 10^{-11}$ & $1.8\times 10^{-9}$  & $2.5\times 10^{-9}$  & $4.5\times 10^{-9}$  \\
    {Single-} & $10^{6}$ & $2.6$  & $1.3\times 10^{-14}$ & $3.3\times 10^{-11}$ & $9.7\times 10^{-10}$ & $1.4\times 10^{-9}$  & $2.4\times 10^{-9}$  \\
    {SMBH   } & $10^{7}$ & $1.0$  & $2.9\times 10^{-16}$ & $7.1\times 10^{-12}$ & $3.7\times 10^{-10}$ & $4.8\times 10^{-10}$ & $8.6\times 10^{-10}$ \\
    {       } & $10^{7}$ & $1.75$ & $2.2\times 10^{-16}$ & $5.3\times 10^{-12}$ & $2.4\times 10^{-10}$ & $3.2\times 10^{-10}$ & $5.6\times 10^{-10}$ \\
    {       } & $10^{7}$ & $2.6$  & $1.3\times 10^{-16}$ & $2.8\times 10^{-12}$ & $1.4\times 10^{-10}$ & $2.1\times 10^{-10}$ & $3.5\times 10^{-10}$ \\
    \hline
    {Dual-  } & $10^{6}$ & $1.75$ & $2.7\times 10^{-10}$ & $3.1\times 10^{-8}$  & $4.6\times 10^{-7}$  & $1.3\times 10^{-6}$  & $1.8\times 10^{-6}$  \\
    {SMBHs  } & $10^{7}$ & $1.75$ & $4.6\times 10^{-11}$ & $2.8\times 10^{-8}$  & $1.0\times 10^{-6}$  & $4.2\times 10^{-6}$  & $5.2\times 10^{-6}$  \\
    \enddata
    \tablecomments{The QPE rates ($\Gamma$) at different period ranges are listed in columns (4)--(8)}.
\end{deluxetable*}

Figure \ref{fig:period_dis} also shows the constitution of QPEs 
grouped by the nature of their progenitor binaries.  
In a dual-SMBH system of $M_{\rm{SMBH}}=10^{6} (10^{7})M_{\odot}$, 
the long-period QPEs are predominately produced by MS-MS binaries, 
with only about $5\ (10)\%$ coming from CO-MS binaries. 
As the periods decrease to $P\lesssim 30 (100)\ \rm{days}$, 
MS-MS binaries suffer severely from binary merger, 
and the contribution from CO-MS binaries gradually gets more prominent. 
The QPEs of the shortest periods with $P\lesssim 3 (10)\ \rm{days}$ 
are produced exclusively from sBH-MS binaries 
because sBHs are more massive than WDs and NSs, 
and the periods of Hills-captured stars decrease 
with the increasing mass of the ejected objects (see Equation (\ref{equ:Pcap})). 
The minimum periods that can be achieved are about $0.6 (2)$ days. 
We note that the critical periods introduced above approximately 
follow $\propto M_{\rm{SMBH}}^{1/2}$, as suggested by Equation (\ref{equ:Pcap}).

\section{Discussions and Conclusions} \label{sec:dis}   

The immediate repeated partial TDEs of Hills-captured MS stars from 
tidally separated binaries 
are suggested to be a promising channel to produce QPEs \citepalias{cufari_using_2022}. 
Based on this model, we calculate the formation rate of QPEs of different periods 
in a single-SMBH system and a dual-SMBH system respectively.  
We show that the QPEs of periods $\leq 10^{4}\ \rm{days}$ are produced at a rate of 
$\Gamma_{\rm{s}}\simeq 10^{-10}$--$10^{-8}\ \rm{yr}^{-1}$ 
in a single-SMBH system of SMBH mass 
$10^{6}$--$10^{7}M_{\odot}$, 
regardless of the shape of the inner stellar density. 
In an equal-mass dual-SMBH system, 
the QPE rate could be 3--4 orders of magnitude higher and reach 
$\Gamma_{\rm{d}}\simeq 10^{-6}$--$10^{-5}\ \rm{yr}^{-1}$ 
when two SMBHs are just about to become self-gravitationally bound.

We note that our calculations do not count 
the number of flares that can be produced by a Hills-captured star, 
which depends on the subsequent evolution of its orbit and inner structure 
after partial TDEs \citep[e.g.,][]{bandopadhyay_repeating_2024,chen_fate_2024}. 
Based on Figure 13 in \citet{chen_fate_2024}, 
we find that the Hills-captured star investigated here is guaranteed to produce 
repeating ($\geq 2$) flares given its tightly bound orbit, 
making it either a QPE with $\geq 3$ flares 
or a twice-repeated TDE 
(with an interval corresponding to the period referred above), 
which could be the case for some repeating transients showing only two flares so far  
\citep[e.g.,][]
{malyali_rebrightening_2023,somalwar_first_2023,wevers_live_2023,lin_unluckiest_2024}. 
Therefore, our results actually show the combined rate of QPEs and twice-repeated TDEs, 
or in other words, the upper rate limit for either of them.

It is suggested that QPEs could also be produced by Hills-captured stars that 
are not partially disrupted immediately 
but inspiral into the SMBH due to gravitational wave (GW) emission (stellar EMRIs) 
and undergo Roche Lobe overflow on mildly eccentric orbits 
\citep[e.g.,][]{linial_unstable_2023,lu_quasiperiodic_2023}. 
We calculate the formation rate of QPEs in this stellar EMRI scenario 
in our fiducial model ($\gamma=1.75$) of single-SMBH systems. 
The formation of a stellar EMRI requires that 
the Hills-captured star is not (partially) disrupted immediately, 
i.e., $r_{\rm{p}}>\mathcal{R}_{\rm{pt}}$, 
and its postcapture orbit evolution is dominated by GW emission 
rather than two-body scatterings, 
where we use Equation (6) in \cite{sari_tidal_2019} as the criterion. 
Accordingly, we count the Hills-captured stars meeting the two requirements above 
and compute the rate with the same procedure 
as introduced in Section \ref{subsec:single-SMBH}.   
The dependence on the period (factor $F_{\rm{P}}$) is ignored as 
the orbit period would shrink significantly during the EMRI stage to $\lesssim$ a few days. 
We find the QPE rate in the stellar EMRI channel to be 
\begin{eqnarray}\label{equ:rate_emri}
    \Gamma_{\rm{s,EMRI}}(M_{\rm{SMBH}}=10^{6}M_{\odot})
    &\approx & 
    1.5\times 10^{-11}\ \rm{yr}^{-1} 
    \left(\frac{f_{\rm{b0}}}{10^{-3}}\right) 
    \left(\frac{f_{\rm{m}}}{0.25}\right)
    \left(\frac{f_{\rm{EMRI}}}{3\times 10^{-4}}\right)
    \left(\frac{\Gamma_{\rm{TDE}}}{2\times 10^{-4}\ \rm{yr}^{-1}}\right) \nonumber \\
    \Gamma_{\rm{s,EMRI}}(M_{\rm{SMBH}}=10^{7}M_{\odot})
    &\approx & 
    4.8\times 10^{-9}\ \rm{yr}^{-1} 
    \left(\frac{f_{\rm{b0}}}{10^{-3}}\right)
    \left(\frac{f_{\rm{m}}}{0.25}\right)
    \left(\frac{f_{\rm{EMRI}}}{0.2}\right)
    \left(\frac{\Gamma_{\rm{TDE}}}{10^{-4}\ \rm{yr}^{-1}}\right),
\end{eqnarray}
where $\Gamma_{\rm{TDE}}$ is the TDE rate for single field stars, 
$f_{\rm{b0}}$ is the fraction of compact binaries with $a_{*}\lesssim 0.2\ \rm{au}$, 
$f_{\rm{m}}$ accounts for the consumption due to binary mergers, 
and $f_{\rm{EMRI}}$ is the fraction of Hills-captured stars undergoing stellar EMRIs 
after the tidal separations of those compact binaries 
numerically evaluated from our results. 
The factor $f_{\rm{EMRI}}$ is significantly smaller when $M_{\rm{SMBH}}=10^{6}M_{\odot}$ 
because the parameter space for stellar EMRIs (see Figure 2 in \citealp{sari_tidal_2019}) 
becomes tiny after considering the consumption due to partial TDEs.  
Previous studies analytically estimate 
$\Gamma_{\rm{s,EMRI}}\approx 10^{-7}$--$10^{-5}\ \rm{yr}^{-1}$, 
assuming that $f_{\rm{b,EMRI}}\simeq 0.1\%$--$10\%$ of binaries 
are compact enough to end up with QPEs 
\citep{linial_unstable_2023,lu_quasiperiodic_2023}. 
Our computed rate is much lower, 
mainly because we adopt a more realistic log-normal distribution of $a_{*}$, 
take binary mergers into consideration, 
and numerically account for the fraction of EMRIs under the competition from partial TDEs, 
leading to a significantly lower 
$f_{\rm{b,EMRI}}(\approx f_{\rm{b0}}f_{\rm{m}}f_{\rm{EMRI}})$.

In observation, 
the formation rate of short-period ($<1\ \rm{day}$) QPEs is 
suggested to be $4\times 10^{-6}(\tau_{\rm{life}}/10\ \rm{yr}) \rm{yr}^{-1}$ per galaxy 
with $\tau_{\rm{life}}$ the unknown QPE lifetime \citep{arcodia_cosmic_2024}. 
By contrast, our computed rate for these short-period QPEs 
in the \citetalias{cufari_using_2022} scenario is extremely lower 
($\lesssim 10^{-14}\ \rm{yr}^{-1}$) even after enhancement in a dual-SMBH system. 
Moreover, the contribution from the stellar EMRI scenario with a rate of 
$10^{-11}-10^{-8}\ \rm{yr}^{-1}$ is still insufficient to explain the observation rate. 
Our results thus suggest that short-period QPEs are unlikely to form 
by Hills-captured stars through either the \citetalias{cufari_using_2022} or the 
stellar EMRI channel 
in a $M_{\rm{SMBH}}\geq 10^{6}M_{\odot}$ SMBH system based on the rate argument. 
The formation rates of short-period QPEs produced by other scenarios, e.g.,  
repeated partial TDEs of a (possibly Hills-captured; \citetalias{cufari_using_2022}) 
WD by an intermediate–massive black hole \citep{king_gsn_2020,king_quasiperiodic_2022}, 
are yet to be explored, 
and we leave them to a future work.

Regarding the QPEs of longer periods, 
the only observational rate available for reference comes from 
\citet{somalwar_first_2023}, who 
roughly constrain the formation rate of QPEs with periods between 0.3--2.7 yr to be 
$10^{-6}$--$10^{-5}\ \rm{yr}^{-1}$ per galaxy 
based solely on a two-flares transient AT2020vdq, 
whose nature is undetermined, as discussed above. 
This observational rate is orders of magnitude higher than our prediction 
in a single-SMBH system 
but is consistent with the predicted rate in a dual-SMBH system.

Given the enhanced QPE formation rate in a dual-SMBH system, 
we then ask what fraction of observed QPEs in a sky survey is expected to 
come from postmerger galaxies hosting dual SMBHs. 
In principle, the fraction ($f_{\rm{d}}$) can be estimated with 
$f_{\rm{d}}=\Gamma_{\rm{d}}\tau_{\rm{d}}/
(\Gamma_{\rm{s}}\tau_{\rm{s}}+\Gamma_{\rm{d}}\tau_{\rm{d}})$, 
where $\tau_{\rm{s}}$ and $\tau_{\rm{d}}$ are, respectively, 
the durations of the galaxy at normal and postmerger phases. 
We assume that the normal phase lasts 
$\tau_{\rm{s}}\simeq 3\times 10^{9}\ \rm{yr}$ 
considering that a galaxy typically experiences $\simeq 3$ mergers 
during a Hubble time $10^{10}\ \rm{yr}$ \citep{oleary_emerge_2021}, 
during which the rate $\Gamma_{\rm{s}}$ keeps constant. 
Moreover, for the postmerger phase, 
we consider that the QPE rate is proportional to the 
full loss-cone flux and thus increases with decreasing $d_{\rm{SMBH}}$ as 
$\Gamma_{\rm{d}}\propto d_{\rm{SMBH}}^{-5/7}$ after 
$d_{\rm{SMBH}}\leq 30(100)r_{\rm{i}}$ 
when $M_{\rm{SMBH}}=10^{6}(10^{7}) M_{\odot}$ \citep{liu_enhanced_2013}, 
and the dual-SMBH system evolves under dynamical friction such that  
\begin{eqnarray} \label{equ:Tdf}
    \tau_{\rm{d}}\simeq T_{\rm{df}}\simeq \frac{1.17}{\ln{\Lambda}}
    \frac{M_{\rm{p}}}{M_{\rm{s}}}P_{\rm{SMBH}} 
    \approx 3.5P_{\rm{SMBH}},
\end{eqnarray}
where $T_{\rm{df}}$ is the dynamical friction timescale of the dual-SMBH system, 
$M_{\rm{p}}$ is the total mass of the primary galaxy within radius $d_{\rm{SMBH}}$, 
$M_{\rm{s}}$ is the mass of the secondary galaxy tidally truncated at Jacobi radius 
$r_{\rm{J}}\approx d_{\rm{SMBH}}/2$ \citep{binney_galactic_2011}, 
and the Coulomb logarithm is 
$\ln{\Lambda}=\ln{(b_{\rm{max}}/b_{\rm{min}})}\approx 1$ after adopting 
$b_{\rm{max}}=d_{\rm{SMBH}}/2$ and $b_{\rm{min}}=r_{\rm{J}}/2$ \citep{just_dynamical_2011}. 
The relations above and the computed QPE rate 
(see Table \ref{tab:rate} when $\gamma=1.75$) result in  
$f_{\rm{d}}\approx 11\% (88\%)$ when $M_{\rm{SMBH}}=10^{6}(10^{7}) M_{\odot}$. 
Therefore, a nonnegligible and even predominant fraction of QPEs are expected to be 
associated with dual SMBHs, 
which explains the preference of finding (long-period) QPEs in postmerger galaxies.

We should mention that the enhanced QPE rate in dual-SMBH systems 
calculated in this work is specifically for two equal-mass SMBHs just about to become bound. 
It is likely that the rate could get further boosted when considering a bound SMBH binary 
interacting with bound stellar binaries in the nuclear stellar cusp, 
especially when the SMBH binary is of very unequal mass after a minor merger 
\citep[e.g.,][]{chen_enhanced_2009,chen_tidal_2011}, 
and we plan to address this topic in the next work.

\begin{acknowledgments}
    We thank the anonymous referee for thoughtful comments and suggestions 
    that helped to improve the manuscript. 
    This work is supported by the National Natural Science Foundation of China (NSFC No.11721303) and
    China Manned Space Project with No. CMS-CSST-2021-A06. 
    S. L. also acknowledges the support by NSFC under grant No.12473017.
\end{acknowledgments}

\software{
NumPy \citep{harris2020array}, 
SciPy \citep{2020SciPy-NMeth}, 
astropy \citep{2022ApJ...935..167A}, 
pandas \citep{mckinney-proc-scipy-2010}, 
Matplotlib \citep{Hunter:2007}, 
IPython \citep{PER-GRA:2007}
}.



\end{document}